\PassOptionsToPackage{hyphens}{url}
\PassOptionsToPackage{hidelinks}{hyperref}
\documentclass[runningheads]{llncs}
\usepackage[T1]{fontenc}

\usepackage[type={CC},modifier={by},version={4.0},imagewidth=5em]{doclicense}
\usepackage[inline]{enumitem}
\usepackage{amsmath}
\usepackage{amssymb}
\usepackage[frozencache]{minted}
\usepackage{listings}
\usepackage{fontawesome}
\usepackage{relsize}
\usepackage{soul}

\usepackage{hyperref}
\usepackage[capitalize]{cleveref}
\usepackage[svgnames]{xcolor}

\usepackage{orcidlink}

\newif\ifdraft\draftfalse
\newif\iflong\longfalse

\ifdraft
\usepackage{todonotes} 
\setuptodonotes{inline} 
\newcommand\pagest[1]{ (#1 pages)}
\pdfpagesattr{/CropBox [92 100 523 740]} 
\else
\usepackage[disable]{todonotes}
\newcommand\pagest[1]{}
\fi

\newcommand\mypara[1]{\vspace{-.5em}\subsubsection{#1}}

\DeclareMathOperator{\dom}{dom}
\DeclareMathOperator{\mday}{day}
\DeclareMathOperator{\mmonth}{month}
\DeclareMathOperator{\myear}{year}
\newcommand\jan{\text{\mintinline{ocaml}{Jan}}}
\newcommand\feb{\text{\mintinline{ocaml}{Feb}}}
\newcommand\mar{\text{\mintinline{ocaml}{Mar}}}
\newcommand\apr{\text{\mintinline{ocaml}{Apr}}}
\newcommand\jun{\text{\mintinline{ocaml}{Jun}}}
\newcommand\sep{\text{\mintinline{ocaml}{Sep}}}
\newcommand\nov{\text{\mintinline{ocaml}{Nov}}}

\usepackage{stmaryrd}
\usepackage{xspace}
\usepackage{xcolor}
\usepackage{mathpartir}

\usepackage{tikz}
\usetikzlibrary{arrows,shapes,positioning,fit,calc,decorations.pathreplacing,shapes.misc,shapes.geometric,decorations.pathmorphing,tikzmark,chains,tikzmark,scopes,matrix,backgrounds,patterns,overlay-beamer-styles,shadows,graphs}

\definecolor{OwlRed}{RGB}{255,92,168}
\definecolor{OwlGreen}{RGB}{90,168,0}
\definecolor{OwlBlue}{RGB}{0,152,233}
\definecolor{OwlYellow}{RGB}{ 242, 147, 24}
\colorlet{OwlCyan}{OwlGreen!50!OwlBlue}
\colorlet{OwlOrange}{OwlRed!50!OwlYellow}
\colorlet{OwlBrown}{OwlRed!50!OwlGreen}
\colorlet{OwlViolet}{OwlRed!50!OwlBlue}

\colorlet{Py}{OwlCyan}
\colorlet{C}{OwlOrange}
\colorlet{Uni}{OwlBlue}

\tikzstyle{dom} = [draw, rectangle, rounded corners, inner sep = 3pt,font=\footnotesize]
\tikzstyle{combiner} = [draw, circle, inner sep = 0pt, minimum size=15pt, font=\footnotesize]
\tikzstyle{iterator} = [dom,fill=MidnightBlue!40]
\tikzstyle{memory} = [dom,fill=JungleGreen!40]
\tikzstyle{scalar} = [dom,fill=Tan!40]
\tikzstyle{numeric} = [dom,fill=Orchid!40]
\tikzstyle{lib} = [dom,fill=BrickRed!40]
\tikzstyle{shareddomain} = [dom,fill=OwlBlue!50]
\tikzstyle{optshareddomain} = [dom,fill=OwlBlue!50]
\tikzstyle{clangdomain} = [dom,fill=OwlOrange!50]
\tikzstyle{optclangdomain} = [dom,fill=OwlOrange!50]
\tikzstyle{pylangdomain} = [dom,fill=OwlCyan!50]
\tikzstyle{analysisdomain} = [dom,fill=OwlRed!50]

\newcommand{\shareddomain}[1]{\tikz[baseline=-0.5ex]\node[shareddomain] {#1};}
\newcommand{\optshareddomain}[1]{\tikz[baseline=-0.5ex]\node[shareddomain] {#1};}

\newcommand{\pylangdomain}[1]{\tikz[baseline=-0.5ex]\node[pylangdomain] {#1};}

\newcommand{\sequence}{\tikz[baseline=-0.5ex]\node[combiner,fill=Coral!20] {$\fatsemi$};\xspace}

\newcommand{\reducedprod}{\tikz[baseline=-0.5ex]\node[combiner,fill=Coral!20] {$\wedge$};\xspace}
\newcommand{\closure}{\overset{*}{\rightarrow}\xspace}

\let\ls\lstinline
\newcommand\fstar{F$^\star$\xspace}
\newcommand\mopsa{Mopsa\xspace}

\newcommand{\fref}[1]{\cref{fig:#1}}
\newcommand{\sref}[1]{\cref{sec:#1}}

\newcommand{\rndup}{\ensuremath{\text{rnd}_\uparrow}\xspace}
\newcommand{\rnddown}{\ensuremath{\text{rnd}_\downarrow}\xspace}
\newcommand{\rnderr}{\ensuremath{\text{rnd}_\bot}\xspace}
\newcommand{\rndarb}{\ensuremath{\text{rnd}_r}\xspace}

\newtheorem{nonprop}{Non-Property}

\newcommand\D{\mathcal{D}}
\newcommand\N{\mathcal{N}}
\renewcommand\P{\mathcal{P}}
\newcommand\V{\mathcal{V}}
\newcommand\Z{\mathbb{Z}}
\newcommand\A[1]{#1^\#}
\newcommand\setst[2]{ \{ #1 \, | \, #2 \}}
\newcommand\set[1]{ \{ #1  \}}
\newcommand\upa{\uparrow}
\newcommand\doa{\downarrow}
\newcommand\E[2][]{\mathbb{E}_{#1} \llbracket #2 \rrbracket}
\renewcommand\S[2][]{\mathbb{S}_{#1} \llbracket #2 \rrbracket}
\newcommand\Val{\text{Val}}
\newcommand\cE{\mathcal{E}}
\newcommand\ud{\updownarrow}

\usepackage{wrapfig}

\crefname{section}{Sec.}{Secs.}
\crefname{figure}{Fig.}{Figs.}
\crefname{example}{Ex.}{Exs.}
\crefname{theorem}{Th.}{Ths.}

\BeforeBeginEnvironment{minted}{\vspace{-.25cm}}
\AfterEndEnvironment{minted}{\vspace{-.5cm}}
\setminted{fontsize=\scriptsize}
\setmintedinline{fontsize=\footnotesize}

\usepackage{enumitem}
\setlist[itemize]{leftmargin=*,noitemsep, topsep=0pt}
\setlist[description]{leftmargin=*,noitemsep, topsep=0pt}

\usepackage[numbers,sort&compress]{natbib}
\title{Formalizing Date Arithmetic and Statically Detecting Ambiguities for the Law}
\titlerunning{Date Arithmetic: Semantics and Analysis}

\author{Raphaël Monat\inst{1}\thanks{Equal contribution}\orcidlink{0000-0001-8487-0326} \and
  Aymeric Fromherz\inst{2}\textsuperscript{\thefootnote}\orcidlink{0000-0003-2642-543X} \and
  Denis Merigoux\inst{2}\orcidlink{0000-0003-2247-0938}}
\authorrunning{R. Monat, A. Fromherz and D. Merigoux}
\institute{
  Univ. Lille, Inria, CNRS, Centrale Lille, UMR 9189 CRIStAL, F-59000 Lille, France\\
  \and Inria Paris, France
  \email{\{raphael.monat,aymeric.fromherz,denis.merigoux\}@inria.fr}}

\begin{document}

\maketitle              
\begin{abstract}
Legal expert systems routinely rely on date computations to determine the eligibility
of a citizen to social benefits or whether an application has been filed
on time. Unfortunately, date arithmetic exhibits many corner cases, which are
handled differently from one library to the other, making faithfully transcribing
the law into code error-prone, and possibly leading to heavy
financial and legal consequences for users.

In this work, we aim to provide a solid foundation for date arithmetic working
on days, months and years. We first present a novel, formal semantics for date
computations, and formally establish several semantic properties through a
mechanization in the \fstar proof assistant. Building upon this semantics, we
then propose a static analysis by abstract interpretation to automatically
detect ambiguities in date computations. We finally integrate our approach in
the Catala language, a recent domain-specific language for formalizing
computational law, and use it to analyze the Catala implementation of the French
housing benefits, leading to the discovery of several date-related ambiguities.

\keywords{Verification, Semantics, Abstract Interpretation}
\end{abstract}

\ifdraft
\linenumbers
\else
\fi 
\section{Introduction}
From filesystems to web servers, time representations are pervasive in modern
computer systems. While several libraries and standards were proposed throughout
the years, current well-established approaches such as Unix time~\cite{unixtime}
used in the standard C library or Windows' FILETIME~\cite{windowstime}
represent dates and time as a number of seconds
or nanoseconds that have elapsed since an arbitrary date.

This approach is sufficient for many usecases, in particular when dates
are only used for logging purposes, or for determining the chronology of two events.
However, it does not permit more complex arithmetic, for instance the addition of
months or years, that span a variable number of days. For these usecases,
mainstream programming languages offer different libraries that adopt
different conventions. For example, Python's \texttt{datetime} module~\cite{pydatetime} forbids
the addition of months, while Java's \texttt{java.time} library~\cite{javatime}
silently rounds invalid dates onto the largest pre-existing date, hiding ambiguous
computations from programmers.

Given the variety of libraries and behaviors across languages, programming with date
arithmetic is thus highly error-prone, and developers' assumptions about how dates
behave might vary from project to project. When developing systems whose correctness
is critical and that heavily depend on date computations, such as expert legal systems that rule
our social and financial lives,
this issue becomes highly concerning.
As an example, consider the following excerpt from Section 121 of the US Internal Revenue
Code~\cite{section121}, which defines the ``Exclusion of gain from sale of principal residence''.

\begin{quote}
  \small
  In the case of a sale or exchange of property by an unmarried individual whose
  spouse is deceased on the date of such sale, paragraph (1) shall be applied by
  substituting “\textdollar 500,000” for “\textdollar 250,000” if such sale
  occurs not later than 2 years after the date of death of such spouse and the
  requirements of paragraph (2)(A) were met immediately before such date of
  death.
\end{quote}

This paragraph differentiates
between two cases, depending on whether a sale occurred \emph{not later than 2 years}
after a given date.
While applying this paragraph is straightforward in most real-world cases,
corner cases raise interesting questions. In particular, when considering
leap years, what should be the result of adding two years to February 29th?
When manually computing taxes, lawyers would be able to detect the ambiguity,
and to reach a decision based on legal precedents. If handled automatically
by a computer however, the computation may be done incorrectly;
computing \ls`February 29 2004 + 2 years` in Java using \texttt{java.time} would
return \ls`February 28 2006`, while performing the same computation using the \texttt{date}
utility from Coreutils returns \ls`March 1 2006`.

Similar computations are pervasive in expert legal systems; the corresponding
regulations rely on them to determine whether a citizen is eligible to social
benefits or a resident for tax purposes. Errors in such systems can have
dramatic consequences; case in point, the incorrect implementation of Louvois,
the former French military payroll system, led to several families either receiving
over-payments that they had to reimburse years later, or incomplete paychecks
totaling a few cents~\cite{louvois}. For such critical software, it is therefore
paramount to provide clear semantics for date computations to avoid 
mistakes based on erroneous assumptions about a library's behavior.
Additionally, such a semantics
can form the basis for further analyses, paving the way for the automated
detection of date-related ambiguities as part of the development process.

Unfortunately, while elegant in theory, a universal semantics for dates and date
arithmetic would not be usable in practice; when possible ambiguities are identified
in law texts, legislators oftentimes extend or modify the law itself to avoid them.
For instance, article 641 of the French civil procedure code~\cite{article641} specifies
that, when adding a positive duration to a date to compute a deadline, the rounding,
if needed, should go down. Such articles often have narrow application scopes;
similar articles in other branches of the law might either leave rounding unspecified,
or adopt a different convention. In the US, date computations when filing motions
are heavily specified, however the complexity and amount of corner
cases led to no less than 27 subsequent notes and amendments to
provide clarifications~\cite{cornellrule6}. Other regulations instead attempt
to escape ambiguities due to month or year additions by reducing such computations
to a nonambiguous number of days. Such regulations heavily vary depending on the country
and the branch of law considered: acts from the Council of European Communities
consider that a month should be treated as 30 days~\cite{eecmonths}, while the
Indian Supreme Court took the opposite approach, enacting that the duration of a month
for customs purposes is variable~\cite{indianmonths}.
To enable their adoption in a variety of contexts, date
libraries therefore require their semantics to be configurable by developers.

The lowest granularity of date arithmetic we focus on is the day level.
Our literature review and communications with lawyers in different countries have indeed shown that this kind of date arithmetic is sufficient for the kind of tax and social benefits computations that are the core application target of Catala.

In this paper, we aim to provide a sound foundation for critical software relying
on date computations, through the following contributions:

\mypara{Formally Capturing Date Computations.}
We first present a formal semantics of date computations (\sref{sems}).
Our formalization relies on a base semantics, which is universal and
does not specify a rounding mode but instead provides facilities to
round on-demand. We leverage these facilities to derive a rounding-specific
semantics for different rounding policies. We mechanize this semantics
in the \fstar proof assistant, and prove several theorems establishing
necessary conditions for, e.g., the monotonicity or associativity of
computations (\sref{mechanized}). As part of this mechanization, we also identify
seemingly intuitive properties that do not hold in practice, and exhibit
counter-examples.

\mypara{Automatically Detecting Date Ambiguities.}
Building on the semantics, we define a notion of \emph{rounding-insensitivity},
which captures that the result of evaluating a program's expression does not depend
on the chosen rounding policy (\sref{sa}).
Aiming to automatically identify possibly harmful ambiguities, we then propose a new static analysis based on abstract interpretation~\cite{cousot1977abstract} targeting this 2-safety hyperproperty.
We implement our analysis in the \mopsa static analyzer~\cite{mopsa,smopsa}.
We show that with relational numerical abstract domains, our analysis enables precise reasoning.
In addition, our implementation provides actionable counter-example hints which will help users understand why a given expression is rounding-sensitive.

\mypara{Contribution to Date Arithmetic Libraries.}
To enable the adoption of this work in existing projects,
we implement an OCaml library abiding by our formal semantics, which exposes
common rounding modes, as well as an option to abort when ambiguous computations
are detected. Our library is standalone and open-source, and easily integrable in OCaml
developments.
We also survey the behavior of mainstream date arithmetic libraries (\cref{sec:rw}),
and provide litmus tests that can be used to easily understand how a library behaves with respect to date rounding.

\mypara{Case Study: Integration in the Catala Language.}
To demonstrate the applicability of our approach in real-world programs,
we replace previous handling of dates in the Catala language~\cite{merigoux2021catala},
a recent domain-specific language for formalizing computational law, by our library.
We also extend the \mopsa~\cite{mopsa,smopsa} static analyzer to support a subset of the Catala language,
enabling us to analyze Catala programs for rounding-insensitivity.
We evaluate our approach against an existing Catala
implementation of the French housing benefits, and automatically identify several
date-related ambiguities in the Catala model. This work is in the process of being
upstreamed in the Catala compiler.

\section{Formalizing Date Arithmetic}
\label{sec:sems}
We start this section by presenting a base semantics for date computations, which does
not explicitly specify a rounding policy to handle ambiguous dates.
Dates expressions are presented in \fref{dateexpr}. Dates values are represented in the
year-month-day format of the standard Gregorian calendar, where each component will be represented as an integer.
We also include a $\bot$ element, which represents an error case.
Date expressions consist of either date values, or of the application of one of the
date operators. Date expressions also contain variables, however their treatment is straightforward and orthogonal to this work; we omit them as well as their associated environment in our presentation. Operators are of two kinds: the addition $+_\delta$ of
$n$ years, months, or days, where $n$ is an integer, and the rounding \rndarb of
a date. Our semantics supports three types of rounding: 
\rndup rounds up the current date to the nearest valid date;
\rnddown rounds down, and \rnderr raises an error if the current date is invalid.
A period is a triple of relative integers, respectively representing the numbers of days, months and years.

\begin{figure}[!t]
\[
\begin{array}{lcrl}
  \text{date unit} & \delta & ::= & y \mid m \mid d \\
  \text{rounding mode} & r & ::= & \uparrow~\mid~\downarrow~\mid \bot \\
  \text{values} & v & ::= & (y, m, d) \mid \bot \\
  \text{expressions} & e & ::= & v \mid e~+_\delta~n \mid \rndarb~e \\
  \text{period} & p & ::= & (n_d, n_m, n_y)
\end{array}
\]
\caption{Date expressions}
\label{fig:dateexpr}
\end{figure}

We now define a formal semantics for evaluating expressions. We start by describing
the semantics of date addition, presented in \fref{datesemsadd}.
To match standard date formats, we start counting at 1 for valid days and months;
to simplify the presentation, we will often represent months using their name instead
of their number (e.g., \jan~instead of 1). 
Our semantics is designed to preserve the following invariant: assuming the date
on the left is initially valid, any non-ambiguous computation will return
a valid date. When the computation is ambiguous, the resulting date
is between the largest smaller and the smallest larger valid date.

Our semantics is defined recursively. Consider for instance the addition
of a number of days $n$. If $n$ is small enough to remain in the same month
and year, we are in the terminal case and the rule \textsc{Add-Days} applies.
The first premise of the rule ensures that the date is initially valid.
It relies on an auxiliary function \ls`nb_days`, omitted for brevity,
which computes the number of days for a month in a given year (e.g.,
31 for January, and 28 or 29 for February depending on the year).
Otherwise, we either add a month (rule \textsc{Add-Days-Over}) or remove a month
(rule \textsc{Add-Days-Under2}) and perform a new addition with an updated number
of days. When the initial date is invalid, we return $\bot$ to avoid propagating
large errors and maintain important properties about date semantics that we
prove in \sref{mechanized}. When composing additions, it might therefore be necessary
to apply rounding operators presented later in this section to avoid $\bot$.
One last point of interest in these semantics is the dissymmetry between
the \textsc{Add-Days-Over} and \textsc{Add-Days-Under-*} rules. Since adding
a number of days is never ambiguous, we wish to ensure that, assuming the initial
date is valid, we never apply the \textsc{Add-Days-Err1} or \textsc{Add-Days-Err2}
rules. To do so, when updating the month or year during day addition, we always go
through an intermediate state corresponding to the first day of the month, which
is always a valid day independently of the month and year. For brevity,
we also omit several redundant error cases, where the current month does not belong
to the interval $[1; 12]$; these cases return $\bot$. Following standard notations,
we will denote the transitive closure of our small-step semantics as $\closure$. 

\begin{figure}[!t]
\[
\begin{array}{c}
\inferrule*[lab=Add-Year]
  {~}
  {(y,m,d) +_y n \rightarrow (y+n, m, d)}

\quad\quad

\inferrule*[lab=Add-Month-Under]
  {m + n < 1}
  {(y,m,d) +_m n \rightarrow (y-1, m, d) +_m (n+12)}

\\\\

\inferrule*[lab=Add-Month]
  {1 \le m + n \le 12}
  {(y, m, d) +_m n \rightarrow (y, m+n, d)}

\quad \quad

\inferrule*[lab=Add-Month-Over]
  {m + n > 12}
  {(y,m,d) +_m n \rightarrow (y+1, m, d) +_m (n - 12)}

\\\\

\inferrule*[lab=Add-Days-Over]
  {1 \le d \le \text{nb\_days}(y, m) \qquad d + n > \text{nb\_days}(y, m)}
  {(y, m, d) +_d n \rightarrow ((y, m, 1) +_m 1) +_d (n - (\text{nb\_days}(y, m) - d) - 1)}

\\\\

\inferrule*[lab=Add-Comp]
  {e \rightarrow e'}
  {e +_\delta n \rightarrow e' +_\delta n}

\quad\quad

\inferrule*[lab=Add-Days-Under1]
  {1 < d \le \text{nb\_days}(y,m) \qquad d + n \le 0}
  {(y, m, d) +_d n \rightarrow (y, m, 1) +_d (d - 1 + n)}

\\\\

\inferrule*[lab=Add-Days-Err1]
  {d < 1}
  {(y, m, d) +_d n \rightarrow \bot}

\quad\quad

\inferrule*[lab=Add-Days-Under2]
  {n + 1 \le 0 \qquad (y, m, 1) +_m (-1) \rightarrow (y', m', d')}
  {(y, m, 1) +_d n \rightarrow (y', m', 1) +_d (n + \text{nb\_days}(y', m')) }

\\\\

\inferrule*[lab=Add-Days-Err2]
  {d > \text{nb\_days}(y, m)}
  {(y, m, d) +_d n \rightarrow \bot}

\quad \quad

\inferrule*[lab=Add-Days]
  {1 \le d \le \text{nb\_days}(y, m) \qquad 1 \le d + n \le \text{nb\_days}(y, m)}
  {(y, m, d) +_d n \rightarrow (y, m, d+n)}

\end{array}
\]
\caption{Semantics for date addition}
\label{fig:datesemsadd}
\end{figure}

The last step is now to define semantics for rounding, shown in \fref{datesemsrnd}.
Compared to additions, the rounding semantics is simpler: if the date is already
valid, any mode of rounding leaves the date unchanged (\textsc{Round-Noop}).
Otherwise, rounding down (\textsc{Round-Down}) returns the last day of the current
month, rounding up (\textsc{Round-Up}) returns the first day of the next month,
while the strict rounding mode (\textsc{Round-Err2}) raises an error. In
all cases, if the day is initially negative, rounding returns $\bot$; we will
prove in \sref{mechanized} that this never happens when starting from a valid
date.

\begin{figure}[!t]
\[
\begin{array}{c}
\\\\
\inferrule*[lab=Round-Err1]
  {d < 1}
  {\rndarb(y, m, d) \rightarrow \bot}

\quad\quad

\inferrule*[lab=Round-Err2]
  {d > \text{nb\_days}(y, m)}
  {\rnderr(y, m, d) \rightarrow \bot}

\quad\quad

\inferrule*[lab=Round-Down]
  {d > \text{nb\_days}(y, m)}
  {\rnddown (y, m, d) \rightarrow (y, m, \text{nb\_days}(y, m))}

\\\\

\inferrule*[lab=Round-Noop]
  {1 \le d \le \text{nb\_days}(y, m)}
  {\rndarb (y, m, d) \rightarrow (y, m, d)}

\quad\quad

\inferrule*[lab=Round-Up]
  {d > \text{nb\_days}(y, m) \qquad (y, m, d) +_m 1 \closure (y', m', d')}
  {\rndup(y, m, d) \rightarrow (y', m', 1)}

\end{array}
\]
\caption{Semantics for date rounding}
\label{fig:datesemsrnd}
\end{figure}

Separating additions and rounding has several benefits. Different use cases
might require different rounding modes, and different ways of adding
days, months, and years. For instance, when adding a period such as
1 year and 10 months, some settings might specify that months should be added
first, or that rounding must be performed after adding months, and again after
adding years; our formal semantics enables this flexibility.

The last remaining step is to define additions not just for individual days, months,
or years, but for composite time periods. Building upon our semantics, we can define
this generically for a rounding mode $r$ as follows, and
avoid the need for users to manually call rounding operators.

\vspace{-.5em}
$$e +_r (y, m, d) ::= \rndarb(((e +_ y y) +_m m)) +_d d$$

One point of interest in our derived forms is that we only apply rounding after performing
addition of years and months. Indeed, adding a year should be equivalent to
adding 12 months. However, if we performed rounding after each operation, adding 1 year and
1 month to \ls`February 29 2020` with the rounding-up mode would return \ls`April 1, 2021`
instead of 
\ls`Mar 29, 2021`. We emphasize that, in cases where this behavior would be expected, defining
derived forms corresponding to this semantics would be straightforward using our base semantics.

Based on this semantics, we can now formally define the notion of an ambiguous date expression in \cref{def:ambiguous}.

\begin{definition}[Ambiguous expression]
  \label{def:ambiguous}
  A date expression $e$ is ambiguous if and only if $\rnderr(e) \closure \bot$.
\end{definition}

Note that this intensional definition of ambiguity is equivalent to stating that the an expression $e$ is ambiguous if and only if rounding $e$ in different modes yields different dates.

While the semantics presented in this section focuses on the core, possibly ambiguous computations, 
our work also includes other non-ambiguous operators (omitted for brevity),
e.g., to retrieve the first or last day of a given month. This allows to encode a variety of patterns,
for instance, the second-to-last day of a month by combining date arithmetic with the ``last day of month''
operator, or to rely on a preprocessing phase if months must be treated as 30 days~\cite{eecmonths}.
Our semantics supports reasoning on computations mixing rounding modes.

\section{Mechanizing Semantics}
\label{sec:mechanized}
Building upon the semantics presented in the previous section,
we now present several properties of interest related to date computations
that we will rely upon when designing a static analysis in \sref{sa}. 
As part of our contributions, we mechanize our semantics, related properties and their proofs
inside the \fstar proof assistant~\cite{mumon}.

\subsection{Semantic properties}

As part of our proof development, we separate semantic properties in two categories:
properties established on the base semantics, valid for all derived forms,
and properties derived on specific rounding modes.
In many cases, proofs on derived forms can be performed efficiently by composing
lemmas on base semantics, thus simplifying the proof effort. During development,
we also encode our OCaml implementation of date computations and corresponding theorems
into qcheck \cite{qcheck},
a QuickCheck~\cite{claessen2000quickcheck} inspired property-based testing framework for OCaml.
We mostly used QuickCheck as a fast sanity check before spending time proving lemmas in \fstar.
In particular, our initial intuition for several of the lemmas and theorems presented was often unreliable, omitting corner cases; we used QuickCheck to gain more confidence in our intuition before moving to \fstar.
This encoding allowed us to automatically find most of the counter-examples presented in \sref{nonprop}.

We start by proving that expressions in our semantics always evaluate to a value (possibly $\bot$),
i.e., reduction is never stuck and it terminates.

\begin{theorem}[Normalization]
For any date $d$, and any integer $n$, there exists a value $v_\delta$ such that
$d +_\delta n \closure v_\delta$.
\end{theorem}

In addition to normalization, a useful property about our semantics is a characterization of valid computations:
when using any of the non-abort rounding modes, an addition starting from a valid date will always return a valid date; the
definition of validity is straightforward, but omitted for brevity.
To prove it, we need the following properties on base semantics, which we prove by induction on the reductions.

\begin{lemma}[Well-formedness of day addition]
\label{lem:wfadd}
For any valid date $d$, any integer $n$, and any value $v$,
$d +_{d} n \closure v \Rightarrow v \neq \bot$.
\end{lemma}

\begin{lemma}[Well-formedness of year/month addition]
\label{lem:wfaddym}
For any valid date $d$, any integer $n$, any value $v$, and
$\delta \in \{y, m\}$, we have $d +_\delta n \closure v \Rightarrow v \neq \bot \wedge \text{\normalfont day\_of}(v) \ge 1$.
\end{lemma}

\begin{lemma}[Well-formedness of rounding]
\label{lem:wfrnd}
For any date $d$ such that $d \neq \bot$, any value $v$, and $r \in \{\uparrow, \downarrow\}$,
we have
$\text{rnd}_r~d \closure v \Rightarrow \text{valid}(v)$.
\end{lemma}

We can now state the following theorem on the derived semantics.

\begin{theorem}[Well-formedness]
For any valid date $d$, any period $p$, any value $v$, and $r \in \{\downarrow, \uparrow\}$,
we have $d +_r p \closure v \Rightarrow \text{valid}(v)$.
\end{theorem}

We now present several theorems related to the monotonicity of the addition in our semantics.
Date comparison is defined in the standard way, as the lexicographical order on $(y, m, d)$.
To simplify the presentation, we lift the comparison operators to operate on date expressions,
defined as the comparison on the values obtained by evaluating the expressions.

\begin{theorem}[Monotonicity]
\label{thm:monotonicity}
For any dates $d_1, d_2$, for any period $p$, for $r \in \{\downarrow, \uparrow\}$,
if $d_1 < d_2$, then $d_1 +_r p \le d_2 +_r p$.
\end{theorem} 

A point of interest in this theorem is the discrepancy between bounds:
while the bound in the premise is strict, the bound in the conclusion is loose.
Unfortunately, a stronger version with strict bounds on both sides does not hold;
for instance, two additions involving rounding down of \ls`April 30` and \ls`April 31`
respectively yield the same result.
To prove this theorem, we again need several intermediate lemmas operating on base semantics. First, we establish an equivalence between
adding years and adding months. We then state and prove several monotonicity properties
on the base semantics.
The proof of Theorem~\ref{thm:monotonicity} follows by direct application of these lemmas.

\begin{lemma}[Equivalence of year and month addition]
\label{lem:equiv}
  For all date $d$, for all integer $n$, $d +_y n = d +_m (12 * n)$.
\end{lemma}

\begin{lemma}[Monotonicity of year and month addition]
For all dates $d_1, d_2$, for any integer $n$, for $\delta \in \{y, m\}$, 
$d_1 < d_2 \Rightarrow d_1 +_\delta n < d_2 +_\delta n$.
\end{lemma}

\begin{lemma}[Monotonicity of day addition]
For all valid dates $d_1, d_2$, for any integer $n$, $d_1 < d_2 \Rightarrow d_1 +_{d} n < d_2 +_{d} n$.
\end{lemma}

\begin{lemma}[Monotonicity of rounding]
\label{lem:rndmono}
For all dates $d_1, d_2$, for $r \in \{\downarrow, \uparrow\}$,
$d_1 < d_2 \Rightarrow \text{rnd}_r(d_1) \le \text{rnd}_r(d_2)$.
\end{lemma}

Finally, we state the following lemma, which guarantees that rounding down will always return
a smaller date than rounding up. Additionally, when the addition is not ambiguous,
the two rounding modes return the same result.

\begin{theorem}[Rounding]
\leavevmode
\begin{enumerate}
\item For all date $d$, for all period $p$, $d +_\downarrow p \le d +_\uparrow p$.
\item For all date $d$, for all period $p$, $d +_\bot p \neq \bot \Rightarrow d +_\downarrow p = d +_\uparrow p = d +_\bot p$.
\end{enumerate}
\end{theorem}

We finally characterize the ambiguity of month addition,
a property that we will need to prove the soundness of the static analysis presented
in \cref{sec:sa}.

\begin{theorem}[Characterization of ambiguous month additions]
\label{thm:characterization}
For all valid date $d$, for all integer $n$, for all value $v$ such that
$d +_m n \closure v$, we have
$\text{\normalfont nb\_days(year\_of}(v), \text{\normalfont month\_of}(v)) < \text{\normalfont day\_of}(v) \Leftrightarrow \rnderr(v) \closure \bot$.
\end{theorem}

\subsection{Non-properties and counter-examples}
\label{sec:nonprop}

We now present several seemingly intuitive and ideally useful properties about date semantics
that do not hold in practice\iflong, exhibiting concrete counter-examples for each of them.
\else.\fi

\begin{nonprop}[Commutativity of addition]
For all date $d$, for all periods $p_1, p_2$, for all $r \in \{\downarrow, \uparrow\}$,
we have $(d +_r p_1) +_r p_2 = (d +_r p_2) +_r p_1$
\end{nonprop}

\iflong For any rounding mode that is not abort, addition is not commutative.\else\fi
Consider the case where
$d =$ March 31, $p_1 =$ 1 day, and $p_2$ = 1 month. When adding $p_1$ first and rounding down,
the addition returns April 30, while the result when adding $p_2$ first will be May 1.
Similar examples exist when rounding up, for instance, by setting $d =$ January 29 2023 \iflong(avoiding leap years)\else\fi,
$p_1 =$ 30 days, and $p_2 =$ 1 month.

\begin{nonprop}[Associativity of addition]
For all date $d$, for all periods $p_1, p_2$, for $r \in \{\downarrow, \uparrow\}$,
we have $(d +_r p_1) +_r p_2 = d +_r (p_1 + p_2)$
\end{nonprop}

\iflong For any rounding mode that is not abort, addition is not associative.\else\fi
Consider the case
where $d =$ March 31, $p_1 =$ 1 month, and $p_2 =$ 1 month. In all rounding modes,
adding $p_1$ followed by $p_2$ will require rounding, ultimately yielding
May 30 or June 1,  while directly adding $p_1 + p_2$
returns May 31\iflong, avoiding rounding.\else .\fi

As the addition being associative and commutative is common among most datatypes,
we emphasize that its invalidity for dates can be a source of confusion for programmers;
common optimizations or rewritings of date computations in a seemingly equivalent
way (e.g., replacing \ls`1 month + 1 month` by \ls`2 months`) can lead to
different outcomes. However, these disparities are exclusively due
to occurrences of rounding in computations.
\iflong In the next section, we\else We \fi thus aim
to help programmers when handling date computations by
proposing a static analysis that automatically detects when rounding
might impact the evaluation of expressions.

\section{A Static Analysis For Rounding-Insensitivity} 
\label{sec:sa}
In this section, we leverage our formal semantics to define a sound static analysis automatically verifying date computations programs. Our goal is to statically detect ambiguous computations, whose result depends on the chosen rounding mode.
Indeed, when writing programs whose specification is the law,
choosing the rounding mode arbitrarily is not a possibility;
this would amount to a legal interpretation that exposes the administration operating
the program to be challenged in court if the rounding mode is unfavorable
to a user.
The cost of bearing the responsibility for making technical regulatory choices for administration
personnel has been documented by \citet{torny:halshs-01249084}.

A naive approach would be to flag any program which contains an ambiguous addition.
However, this solution can be overly restrictive: computations can be ambiguous while having no impact on the final outcome of the program.
Consider for example the expression \mintinline{c}{d + 1 month <= March 15 2023}.
If no rounding happens when adding \mintinline{c}{d} and \mintinline{c}{1 month}, then the expression is obviously safe.
Otherwise, we notice that the rounding may only happen to yield the last day of a month, or the next day of the upcoming month.
In both cases, comparing this result with a date in the middle of a month is thus safe.
Instead, we consider a more interesting property called \textit{rounding-insensitivity}, capturing that the evaluation of an expression is the same for both rounding modes.

At a high-level, our analysis works by tracking constraints over the day, month, and year of a date, through the YMD domain (\cref{sec:ymd}).
The YMD domain is fully parametric in a numerical abstract domain, and works by translating date constraints into numerical constraints.
We discuss the choice of numerical abstract domains in \cref{sec:num}, in order to obtain the best precision in the presence of linear constraints and unconstrained dates.
We analyze the computations with both rounding modes and compare the result to decide rounding-insensitivity, which is a 2-safety hyperproperty.
We explain how we lift the YMD domain to these double computations in \cref{sec:double}.
We implemented our analysis within the \mopsa static analysis platform~\cite{mopsa,smopsa}, described in \cref{sec:mopsa}.
\iflong An important design goal of our analysis is that its results should be understandable by law experts, as they will be required to interpret or clarify corresponding law texts when ambiguities are detected.
We rely heavily on disjunctions and relational constraints to maintain the best precision and yield meaningful counter-examples hints; we detail our counter-example generation in \sref{counter-examples}.
\else
We have taken special care in ensuring that actionable counter-examples can be generated in \sref{counter-examples}, paving the way for use by non-experts.
\fi  

We think that abstract interpretation hits a sweet spot to perform this analysis.
Its full automation makes it usable by non-specialists, especially with the provided counter-example hints.
It allows to derive tailored approximations thanks to \cref{thm:characterization}.
The current definition of date addition is recursive and there are non-linear arithmetic constraints involved, which does not work well with SMT.

We use as a motivating example the program shown in \cref{lst:motivating}.
This program has been extracted from a Catala code snippet used to formalize the French housing benefits \cite[Sec. 3.1]{merigoux:hal-03712130}.
We will provide more details on Catala and the extraction to date programs in \cref{sec:catala}.
In this program, we pick two arbitrary, unconstrained dates, perform a date-duration addition of two years, and project the resulting date onto the first day of its month.
The assertion at line 5 expresses the rounding-insensitivity of the comparison between an arbitrary, unconstrained date and the computed date.\footnote{
  Here \texttt{sync(current < limit)} could be reduced to \texttt{sync(limit)}.
  However our analysis will not need it, and will be able to provide counter-example hints also targeting the values of \texttt{current}, improving readability of the output.
}
 The \texttt{sync} predicate, formally defined in \cref{sec:double}, holds if and only if the evaluation of its expression in both rounding modes yields the same result, meaning that the expression is rounding-insensitive.

\begin{figure}[!t]
  \begin{minted}[xleftmargin=2em,autogobble,numbers=left]{c}
date current = random_date();
date birthday = random_date();
date intermediate = birthday + [2 years, 0 months, 0 days];
date limit = first_day_of(intermediate);
assert(sync(current < limit));
  \end{minted}
  \caption{Example extracted from Catala code modeling the French housing benefits}
  \label{lst:motivating}
\end{figure}

The programs we consider in this section are written in a standard, toy imperative language.

\subsection{The YMD domain combinator}
\label{sec:ymd}

The YMD domain translates constraints on the year, month and day of a date into numerical constraints over three integer variables.
These numerical constraints are handled by a numerical abstract domain, described in \cref{def:numabs}.
The YMD domain can be seen as a domain combinator, or a functor relying on a numerical abstract domain -- we will discuss the chosen instantiation in \cref{sec:num}.
This domain works at a fixed rounding mode.

\begin{definition}[Numerical abstract domain]
  \label{def:numabs}
  In the following, a numerical abstract domain is a lattice $\A\N$ on which the
  following operations are defined:
  \begin{itemize}
    \item The assignment, \texttt{assign}, between a variable and an expression in a given abstract environment yields another abstract environment.
    \item The boolean filtering of a state, \texttt{assume}, filters an abstract environment to enforce that a boolean expression holds.
  \end{itemize}
  This domain is further defined by a concretization function
  $\gamma_N : \A\N \to \P(\V \to \Z)$ mapping numerical abstract environments to a set of concrete integer environments it represents.
  We assume the numerical abstract domain is sound.
\end{definition}

Given a date variable $v$, the YMD domain will create new auxiliary (or ghost) variables $\myear(v),\allowbreak \mmonth(v), \mday(v)$, which do not exist in the original program but simplify reasoning. This is an approach we borrow from the deductive
verification community, and that has been used in static analyses both in the work
of \citet{DBLP:conf/vmcai/ChevalierF20} as well as in \mopsa.

We provide a formal definition of the concretization, which defines the meaning of the YMD domain, and illustrate it on an example.

\begin{definition}[Concretization of the YMD domain]
  The concretization of the YMD domain is formally defined in \cref{fig:concr:ymd}.
  It explains how an abstract numerical environment  $\A n \in \A\N$ can be interpreted into a set of date environments $e \in \V \to \D$ mapping variables to dates.
  To construct these date environments, we first pick an integer environment $\rho \in \V \to \Z$ from the concretization of the numerical abstract domain $\gamma_N(\A n)$.
  The date environments will have as domain definition the date domain of function $\rho$,  $\normalfont \text{dates\_dom}(\rho)$, which is the set of variables where auxiliary year, month and day variables are defined in $\rho$.
  For each of those variables $v \in \normalfont \text{dates\_dom}(\rho)$, $e(v)$
  corresponds to the date defined by the auxiliary variables in $\rho$, provided
  that the date is valid.
\end{definition}

\begin{figure}[!t]
  \begin{align*}
     & \text{dates\_dom} : \left \{
    \begin{array}{ccl}
      (\V \to \Z) & \to     & \P(\V)                                                       \\
      \rho        & \mapsto & \setst{ v }{\myear(v), \mmonth(v), \mday(v) \in \dom(\rho) }
    \end{array}
    \right .                                        \\
     & \gamma_{\text{\footnotesize YMD}} : \left \{
    \begin{array}{ccl}
      \A\N & \to     & \P(\V \to \D)                                                                                                                        \\
      \A n & \mapsto & \bigcup_{ \rho \in \gamma_{\N}(\A n) } \setst{ e }{ \dom(e) = \text{dates\_dom}(\rho) \wedge \forall v \in \dom(e), e(v) = (y, m, d) \\[.25em]
           &         & \wedge \text{valid}(y, m, d) \wedge y = \rho(\myear(v)) \wedge m = \rho(\mmonth(v)) \wedge d = \rho(\mday(v)) }
    \end{array} \right .
  \end{align*}
  \caption{Concretization of the YMD domain}
  \label{fig:concr:ymd}
\end{figure}

\begin{example}[Concretization]
  Let us assume our numerical domain is a map from variables to intervals, and consists of the following state: $\A n = \mday(d) \in [1, 31] \wedge \mmonth(d) \in [1, 12] \wedge \myear(d) = 2023$.
  In that case, the concretization is the set of date environments $e$ defined on variable $d$ such that $e(d)$ can be any valid date of 2023.
  Thus, there is a date environment $e \in \gamma_{\text{\footnotesize YMD}}(\A n)$ such that $e(d) = (2023, 1, 31)$.
  However, there is no date environment such that $e(d) = (2023, 2, 29)$ and $e \in \gamma_{\text{\footnotesize YMD}}(\A n)$ because the date is invalid (2023 is not a leap year).
\end{example}

The YMD domain handles the following transfer functions:
\begin{itemize}
  \item Accessors to the day, month or year number of a date.
        Given a date encoded as a variable $v$, these functions return the associated variable $\mday(v), \mmonth(v),\allowbreak \myear(v)$ respectively.
  \item Projection of a date on the first day of the month: given a date encoded as a variable $v$, this function creates a new date having the same auxiliary month and year variables. The day auxiliary variable is set to 1.
        A similar operator working on the last day of the month can be defined.
  \item The main part of the YMD domain is the transfer function handling month addition and potential rounding originating from this addition. We define it below, argue it is sound, and illustrate it on an example (in \cref{sec:num}).
        As we have proved in \cref{lem:equiv}, additions on years and months can be reduced to additions on months.
        Our current, potentially ambiguous, real-world examples taken from legislative code do not use day addition;
        as it is never ambiguous, we thus do not currently implement it.
        Given its similarity to month addition, we do not foresee any technical difficulty doing so.
  \item The YMD domain also provides a transfer function to compare two dates.
        It is induced by the lexicographic definition of concrete date comparisons and partitions the results to improve the precision.
\end{itemize}

\mypara{Transfer function for month addition.}
We provide a simplified OCaml implementation for the month addition transfer function in \cref{fig:month:add}.
The transfer function takes as parameter a \texttt{date}, represented as a variable; a concrete number of months; an input abstract state; and a rounding mode chosen for date computations.
It will return a case disjunction\footnote{These disjunctions can be seen as a partitioning of the abstract state. In this section we consider everything is partitioned to improve the precision. Our implementation supports limiting the number of partitions.}
  of type \texttt{cases}: a list of \texttt{case}, each consisting in an expression and an abstract state.
We start by defining \texttt{day, month, year}, which are expressions representing the day, month and year number of \texttt{date} through auxiliary variables.
The resulting month and year are computed through non-linear expressions.
Similarly to the semantics, we encode months as integers to perform arithmetic operations, and start our numbering at 1 for January.
The transfer function performs a case disjunction to detect if date rounding will happen, following the characterization of ambiguous month additions (\cref{thm:characterization}).
This case disjunction checks whether the day of the date is compatible with the number of days in the resulting month (and year, as February has one more day during leap years).
This disjunction is encoded thanks to the \texttt{switch} utility, which takes as input an abstract state and a list of tuple of expressions and continuations.
Given a tuple $\texttt{(cond, k)}$, the input abstract state is filtered to satisfy the expression \texttt{cond} (by delegation to the numerical abstract domain).
The resulting abstract state is fed to the continuation, which yields a case.
The cases we encounter during the addition are:
\begin{description}
  \item[Rounding to 29 Feb. of a leap year.]
    If the resulting month is February of a leap year, and the current day number is greater than 29, we will have to perform date rounding.
    We do so using the auxiliary \texttt{round} function.
    Depending on the rounding mode, it either chooses the provided date, or the first of the month afterwards.
    This date is then returned in its corresponding abstract state using \texttt{mk\_date}, whose implementation is not detailed.
  \item[Rounding to 28 Feb. of a non-leap year.]
    \iflong If the resulting month is February of a non-leap year and the current day number is greater than 28,
      we apply rounding as in the previous case.
    \else
      Similar case omitted for brevity.
    \fi 
      
  \item[Rounding to a 30-day month.]
    If the current day number is 31 but the resulting month has 30 days (i.e, it is either April, June, September or November), we also have to perform a rounding, either to the 30th of the resulting month, or the 1st of the month after.
  \item[Other cases.]
    No rounding happens, the day number remains the same.
\end{description}

\begin{figure}[!t]
  \begin{minted}[xleftmargin=2em,numbers=left]{ocaml}
type case = expr * state
type cases = case list

let switch abs = List.map (fun (cond : expr, k : state -> case) -> k (assume cond abs))

let is_leap (y : expr) : expr = (y % 4 = 0 && y % 100 <> 0) || (y % 400 = 0)

let round (r : rounding) (d m y : expr) (abs : state) : case =
  match r with
  | RoundDown ->
    mk_date d m y abs
  | RoundUp ->
    let succ_m = 1 + res_month % 12 in
    let succ_y = y + res_month / 12 in
    mk_date 1 succ_m succ_y abs

let add_months (r : rounding) (date : var) (nb_m : int) (abs : state) : cases =
  let day = day_of date in
  let month = month_of date in
  let year = year_of date in
  let res_month = 1 + (month - 1 + nb_m) % 12 in
  let res_year = year + (month - 1 + nb_m) / 12 in
  switch abs
  [
    (* Rounding to 29 Feb. of a leap year *)
    day > 29 && res_month = Feb && is_leap res_year, round r 29 res_month res_year;
    (* Rounding to 28 Feb. of a non-leap year *)
    day > 28 && res_month = Feb && not (is_leap res_year), round r 28 res_month res_year;
    (* Rounding to a 30-day month *)
    day > 30 && is_one_of res_month [Apr;Jun;Sep;Nov], round r 30 res_month res_year;
    (* No rounding *)
    mk_true, mk_date day res_month res_year
  ]

let dates_lt (d1 d2 : var) (abs : state) : cases =
  switch abs
  [
    (year_of d1) < (year_of d2), mk_true;
    (year_of d1) > (year_of d2), mk_false;
    (year_of d1) = (year_of d2) && (month_of d1 < month_of d2), mk_true;
    (year_of d1) = (year_of d2) && (month_of d1 > month_of d2), mk_false;
    (year_of d1) = (year_of d2) && (month_of d1 = month_of d2)
      && (day_of d1 < day_of d2), mk_true;
    (year_of d1) = (year_of d2) && (month_of d1 = month_of d2)
      && (day_of d1 >= day_of d2), mk_false;
  ]
\end{minted}
  \caption{Abstract transfer functions for month addition and date comparison}
  \label{fig:month:add}
\end{figure}

Note that \texttt{add\_months}, \texttt{round} and \texttt{is\_leap} define syntactic expressions, which will be delegated through \texttt{assign} and \texttt{assume} to the numerical abstract domain.
The expressions at lines 6, 13, 14, 21--22, 26, 28, 30 are not directly evaluated: they will be interpreted by the \texttt{assume} of the numerical abstract domain during the evaluation of the \texttt{switch} function.
The definition of the transfer function for month addition assumes the number of months to add is known as a concrete integer.
This is not restrictive in practice: all programs we extracted from Catala in \cref{sec:catala} only perform date-month addition with a concrete number of months.

The proof of soundness of the abstract month addition, is not formalized in \fstar and omitted for brevity.
However, it is a direct application of the characterization of ambiguous month additions established in \cref{thm:characterization}, and proved formally in \fstar.

The analysis may refine constraints on a day, month or year auxiliary variable.
These constraints could then entail new constraints on other auxiliary variables of the same date to represent only valid dates.
This propagation phase is performed by the strengthening operator described below, which is sound as it only removes invalid dates, which are not taken into account by the concretization.

\mypara{Strengthening operator.}
The strengthening operator enforces the following:
\begin{itemize}
  \item If the month is February, the day is less than 30.
  \item If the month is April, June, September of November, the day is less than 31.
  \item If the date is February 29, we know the current year is a leap year.
    We enforce that the year number is divisible by 4, which is a necessary condition.
    \iflong \footnote{This is not sufficient, as years divisible by 4 and 100, but not by 400 are not leap. Thus 1900 and 2100 are not leap years, but 2000 is.}\else\fi
\end{itemize}

\mypara{Comparison transfer function.}
The transfer function for date comparisons is \mintinline{ocaml}{dates_lt} in \cref{fig:month:add}; it encodes a lexicographic comparison.

\subsection{Instantiating YMD with a combination of numerical domains}
\label{sec:num}

The YMD domain is fully generic in the numerical abstract domain it relies on to translate date constraints into constraints over integers.
We describe how we chose a combination of numerical abstract domains to get the best precision possible in the presence of non-linearity and unconstrained dates.

We initially started using intervals and congruences for our first tests.
Due to its convexity, the interval domain was unable to precisely represent months where the day number may be rounded to 30 days during the date-month addition (line 30 of \cref{fig:month:add}).
Thus, we added a domain of powerset of integers (of size at most 4) to be precise enough for this usecase.
When \texttt{month} is not a constant, the congruence domain will be unable to precisely represent the resulting month (line 21 of \cref{fig:month:add}), and refine the potential values of \texttt{month} given constraints on \texttt{res\_month}.
This situation happens often in our evaluation; it is shown in our motivating example.
We resolved this precision issue by switching from the congruence domain to the relational, linear congruence domain \cite{pplgrids}.
We also added the polyhedra domain \cite{CousotH78} to keep track of equalities between different day, month and year variables, which happens during analyses on programs with unconstrained dates, as we will show in the upcoming examples.

Our current numerical abstract domain is a reduced product between grids, polyhedra, intervals, and a bounded powerset of integers.
The relational domains rely on the Apron library~\cite{apron}.
The approximation of non-linear computations is performed through linearization techniques \cite{DBLP:conf/vmcai/Mine06}.

\begin{example}
  \label{ex:date:add:1m}

  Let us consider the program below picking an arbitrary, unconstrained date \texttt{d} and then adding one month to \texttt{d}.
  We illustrate the different cases of the transfer function \texttt{add\_months} in this case, assuming we round down.
  \[ \text{\mintinline{c}{date d = random_date();     date d2 = d + [0 years, 1 months, 0 days];}} \]
  \begin{description}
    \item[Rounding to 29 Feb. of a leap year.] In the first case of the transfer function, the numerical domain is able to deduce from the expression \mintinline{ocaml}{day > 29 &&}\allowbreak\mintinline{ocaml}{res_month = Feb} that the day of \texttt{d} is either 30 or 31, and the month is January.
      In the rounding down mode, \texttt{d2} is thus February 29th.
      The relational domain additionally expresses that $\myear(d) = \myear(d2)$.
    \item[Rounding to 28 Feb. of a non-leap year.] Similar case, omitted for brevity.
    \item[Rounding to a 30-day month.]
      The numerical abstract domain infers that \texttt{d} represents the 31st of March, May, August or October, tracked thanks to the bounded set of integers domain.
      As we round down, we deduce that the day of \texttt{d2} is 30, and $\mmonth(\texttt{d}) \in \set{{\apr,\jun,\sep,\nov}}$.
      In that case, the relational domain can also infer that $\myear(d) = \myear(d2)$, as \mintinline{ocaml}{m / 12} will always be zero\iflong\footnote{There is currently an imprecision in our implementation meaning that we can only infer that $\myear(\texttt{d2})$ is $\myear(\texttt{d})$ or $\myear(\texttt{d})+1$. We believe this imprecision comes from the linearization pass used by relational abstract domains and have opened a discussion with Apron's maintainers on this topic.}\else\fi.
    \item[Other cases.] In the last case, the intervals and powerset domains cannot express interesting constraints on \texttt{d} and \texttt{d2}.
      The relational domains are however able to capture key relations:
      \begin{itemize}
        \item The day does not change as there is no rounding: $\mday(\texttt{d}) = \mday(\texttt{d2})$.
        \item Thanks to the grids domain \cite{pplgrids} we can infer linear relations modulo a constant, and thus that the month of \texttt{d2} is the month after \texttt{d}, even if the year changes: $\mmonth(\texttt{d2}) \equiv_{12} \mmonth(\texttt{d}) + 1$, where $\equiv_{12}$ denotes congruence modulo 12.
              Note that since $\mmonth(\texttt{d2})$ is not a constant, the non-relational congruence domain is not sufficient to express this relation.
        \item The year number may be the same, or the successor provided that the month of \texttt{d} is December. We lose a bit of precision, as the last month always creates a year increase in the concrete.\\
              {\small $ 12\myear(\texttt{d}) + \mmonth(\texttt{d}) \leq 12\myear(\texttt{d2}) + 11 \wedge 12\myear(\texttt{d2}) \leq 12\myear(\texttt{d}) + \mmonth(\texttt{d}) + 1$}
      \end{itemize}
  \end{description}
\end{example}

\begin{example}[Addition and strengthening]
  \label{ex:running:add}
  We use our running example from \cref{lst:motivating}, and show what the date addition and the strengthening operator yield for dates \texttt{birthday} and \texttt{intermediate}.
  In this example, we assume the dates are rounded up.
  As we add two years to \texttt{birthday}, two of the four cases described in the month addition previously presented will not apply; we omit them below.
  \begin{description}
    \item[Rounding to 28 Feb. of a non-leap year.] In that case, \texttt{birthday} is a Feb. 29th, and \texttt{intermediate} rounds up to March 1st. We additionally know that $\myear(\texttt{birthday}) + 2 = \myear(\texttt{intermediate})$. The strengthening ensures that $\myear(\texttt{birthday})$ is divisible by 4.
    \item[No rounding.] The day and month numbers of \texttt{birthday} and \texttt{intermediate} are equal. The year condition is similar to the one provided in \cref{ex:date:add:1m}.
  \end{description}
\end{example}

\begin{example}[Comparison]
  \label{ex:running:comp}
  Let us continue with our running example, assuming we are focusing on the partition where \texttt{intermediate} has been rounded up to March 1st (as shown in \cref{ex:running:add}).
  In that case, \texttt{limit} is equal to \texttt{intermediate}.
  Assuming the comparison \texttt{current < limit} holds, we have three different cases, described by the line number in \cref{fig:month:add}. Line 38 yields $\myear(\texttt{current}) < \myear(\texttt{limit})$. Line 40 enforces $\myear(\texttt{current}) = \myear(\texttt{limit}), \mmonth(\texttt{current}) < \mmonth(\texttt{limit})$, so
      $\mmonth(\texttt{current}) \in \set{\text{\jan}, \text{\feb}}$. Line 42 yields that the year and month numbers of \texttt{current} and \texttt{limit} are the same and  $\mday(\texttt{current}) < \mday(\texttt{limit})$.
      This last case is impossible given that $1 \leq \mday(\texttt{current}) \allowbreak \leq 31$ and $\mday(\texttt{limit}) = 1$.
\end{example}

\subsection{Lifting to both rounding modes}
\label{sec:double}

The YMD domain operates at a given, fixed rounding mode.
In this section, we leverage the YMD domain to perform date computations in both rounding modes and thus prove rounding-insensitivity.
This lifting is inspired by \citet{endianness-david}, who analyze product programs to prove endianness portability of C programs.
Here, we keep the product of programs implicit: only the rounding mode changes
between the two executions we will consider.

We start by explaining how the concrete semantics are lifted from a single rounding mode to both.
We assume we have a semantics of expressions  (respectively statements)
$\E[r]{expr}$ (resp. $\S[r]{stmt}$) parameterized by a date rounding mode $r \in
  \set{ \upa, \doa }$.
They take as input sets of environments ($\cE = \V \to \Val$) mapping variables to values (which are either integers or dates), and produce values (resp. environments).

\[ \E[r]{expr} : \P(\cE) \to \P(\Val) \qquad \qquad \S[r]{stmt} : \P(\cE) \to \P(\cE) \]

We define in \cref{fig:sem:bi} the concrete semantics evaluating expressions and statements over both rounding modes, written respectively $\E[\ud]{expr}$ and $\S[\ud]{stmt}$.
We do not delve into the details of product programs, which are defined in the work of \citet{endianness-david}.
In this semantics, the state is now duplicated: $\D = \cE \times \cE$.
We ensure that random operations return the same value in both rounding modes, to avoid spurious desynchronizations.
The sync predicate returns true if and only if the expression evaluates to the same values in both
rounding modes, capturing the rounding-insensitivity of the contained expression.
We use it in the programs we analyze to target the expressions we want to check, as we have already seen in \cref{lst:motivating}.
The evaluation of other expressions is performed pointwise on both rounding modes, and similarly for the assignments.

\begin{definition}
  An expression $e$ is \emph{rounding-insensitive} in a state $d$ if and only if $\E[\ud]{sync(e)}(\{ d \}) = \{ (true, true) \}$.
  This property is encoded in programs by the statement \texttt{assert(sync(e))}.
\end{definition}

\begin{figure}[!t]
  \begin{align*}
     & \E[\ud]{expr} : \P(\D) \to \P(\Val^2)                                                                                                                                                         \\
     & \E[\ud]{\text{random\_date()}}(D) = \setst{ (d, d) }{ d \in \Z^3, valid(d) }                                                                                                                  \\
     & \E[\ud]{\text{sync}(e)}(D) = \bigcup_{(\rho_\upa, \rho_\doa) \in \D} \setst{ (b_u == b_d, b_u == b_d) }{ (b_u, b_d) = \E[\ud]{e}(\rho_\upa, \rho_\doa) }                                      \\
     & \E[\ud]{expr}(D) = \bigcup_{(\rho_\upa, \rho_\doa) \in \D} \setst{ (v_\upa, v_\doa) }{ v_\upa =  \E[\upa]{e}\rho_\upa, v_\doa = \E[\doa]{e}\rho_\doa) }                                       \\
     & \S[\ud]{stmt} : \P(\D) \to \P(\D)                                                                                                                                                             \\
     & \S[\ud]{v = e}(D) = \bigcup_{(\rho_\upa, \rho_\doa) \in D} \set{ (\S[\upa]{v = v_\upa }\rho_\upa, \S[\doa]{v = v_\doa}\rho_\doa), (v_\upa, v_\doa) \in \E[\ud]{e}\set{\rho_\upa, \rho_\doa} }
  \end{align*}
  \caption{Concrete semantics over double evaluation of rounding modes}
  \label{fig:sem:bi}
\end{figure}

\newcommand\xup{\mathrel{\mkern+1mu}\upa\mathrel{\mkern-6mu}}
\newcommand\xdo{\mathrel{\mkern+1mu}\doa\mathrel{\mkern-6mu}}

The abstract semantics mimics the concrete behavior, but works on a single abstract state instead of a set of concrete double states.
The double state is represented by duplicating variables according to their rounding mode in the numerical abstract domain.
A variable \texttt{x} is thus written $\xup \texttt{x}$ (resp. $\xdo \texttt{x}$) to represent the variable when the upper (resp. lower) rounding mode is used.
This duplication is performed in a shallow fashion to improve usability: when performing an assignment \texttt{x = e}, if \texttt{e} evaluates into the same value in both rounding modes, the variable \texttt{x} will not be duplicated into the numerical abstract domain.

\begin{example}[Rounding-sensitivity of the comparison]
  \label{ex:running:rnd}
  Back to our running example, we have shown so far how the YMD domain analyzes the program when rounding up (\cref{ex:running:comp}).
  Continuing with the same relational abstract domain, we show part of the abstract state in the partition focusing on rounding to Feb. 28 of a non-leap year in \cref{eq:abs}.
  In the rounding mode down, \texttt{intermediate} rounds to Feb. 28, and thus \texttt{limit} rounds down to Feb. 1st.
    {\small \begin{equation}
        \label{eq:abs}
        \begin{aligned}
           & \mday(\texttt{current}) \in [1, 31], \mmonth(\texttt{current}) \in [1, 12], \myear(\texttt{current}) \in [-\infty, +\infty] \\
           & \mday(\texttt{birthday}) = 29, \mmonth(\texttt{birthday}) = \text{\feb}, \myear(\texttt{birthday}) \equiv_{4} 0             \\
           & \xup\mday(\texttt{intermediate}) = 1, \xup\mmonth(\texttt{intermediate}) = \text{\mar}                                      \\
           & \xdo\mday(\texttt{intermediate}) = 28, \xdo\mmonth(\texttt{intermediate}) = \text{\feb}                                     \\
           & \xdo\myear(\texttt{intermediate}) = \xup\myear(\texttt{intermediate}) = \myear(\texttt{birthday}) + 2                       \\
           & \xup\mday(\texttt{limit}) = 1, \xup\mmonth(\texttt{limit}) = \text{\mar}
          \xdo\mday(\texttt{limit}) = 1, \xdo\mmonth(\texttt{limit}) = \text{\feb}                                                       \\
           & \xdo\myear(\texttt{limit}) = \xup\myear(\texttt{limit}) = \myear(\texttt{birthday}) + 2
        \end{aligned}
      \end{equation}}

  We exhibit an abstract state where we cannot prove that the expression \texttt{current < limit} is rounding-insensitive.
  The static analysis will consider all cases in the comparison and the evaluation in both rounding modes.
  For the sake of presentation here, we only highlight one case.
  The date comparison operator between \texttt{current} and the rounded up version of \texttt{limit} yields a partition where the years are the same and the month number is less.
  This partition refines the abstract state above with the following constraints:
  {\small \begin{equation}
    \label{eq:const1}
    \begin{aligned}
      \myear(\texttt{current}) = \xup\myear(\texttt{limit})\,
      \wedge
      \xup\mmonth(\texttt{limit}) < \mmonth(\texttt{current}) = \text{\mar}
    \end{aligned}
  \end{equation}}

  Let us now consider the case where the comparison with the rounded down version of \texttt{limit} does not hold,
  when the years and months are the same but the days are not.
  We get the following additional constraints:
  {\small \begin{equation}
    \label{eq:const2}
    \begin{aligned}
       & \myear(\texttt{current}) = \xdo\myear(\texttt{limit})\, \wedge \mmonth(\texttt{current}) = \xdo\mmonth(\texttt{limit}) = \text{\feb}\, \wedge \\
       & \mday(\texttt{current}) \geq \xdo\mday(\texttt{limit}) = 1
    \end{aligned}
  \end{equation}}

  Combining the constraints from \cref{eq:const1,eq:const2} on the abstract state from \cref{eq:abs} gives the following result on \texttt{current}:
  {\small \begin{equation}
    \label{eq:result}
    \begin{aligned}
      \myear(\texttt{current}) = \myear(\texttt{birthday}) + 2 \wedge \mmonth(\texttt{current}) = \text{\feb}
    \end{aligned}
  \end{equation}}

  To summarize, our analysis has been unable to prove the rounding-insensitivity of the expression \texttt{current < limit}, in particular in the case of the abstract state presented in \cref{eq:abs}, refined with constraints from \cref{eq:const1,eq:const2}.
  Thanks to partitioning and relational abstract domains, we know that the proof fails when \texttt{birthday} is a Feb. 29th (of a year $y$ which is divisible by 4, a sound but not complete way to express it is leap).
  In that case, \texttt{intermediate} will either be Feb. 28th or March 1st of $y+2$.
  This entails that \texttt{limit} will either be Feb. 1st or March 1st of $y+2$.
  In the cases where \texttt{current} is a day of February of $y+2$ (\cref{eq:result}), the comparison will effectively be rounding-sensitive.

  The original program did not contain any constraints on \texttt{birthday} or \texttt{current}.
  Note that if we add in the program that the day of \texttt{birthday} is less than 28, our analysis is able to automatically prove the program to be rounding-insensitive.
\end{example}

\subsection{Implementation}
\label{sec:mopsa}

We implemented our approach in the \mopsa static analysis platform~\cite{mopsa,smopsa}.
\mopsa is able to analyze C, Python and multilanguage Python/C programs~\cite{DBLP:conf/sas/OuadjaoutM20,DBLP:conf/ecoop/MonatOM19,DBLP:conf/sas/MonatOM21}, to prove the absence of runtime errors, and to perform portability analysis of C programs~\cite{endianness-david}.
We modified the front-end of a toy imperative language also available in \mopsa to analyze programs performing date arithmetic.
We chose to extend this language for our analysis as we do not require advanced features from C nor Python.
Thanks to \mopsa{}'s modular architecture, we have been able to reuse iterators for intraprocedural analysis
with little code changes.
\begin{wrapfigure}[10]{l}{0.5\textwidth}
  \vspace{-2em}
  \resizebox{.5\textwidth}{!}{
    \begin{tikzpicture}
      \matrix[matrix of nodes] (top) {
\pylangdomain{D.bidates} & \sequence &
\shareddomain{U.program} & \sequence &
\shareddomain{U.intraproc} & \sequence &
\shareddomain{U.ymd} & \sequence &
\\
};

\node[below=3pt of top, rounded corners, fill=gray!10, inner sep=5pt] (bottom) {
\begin{tikzpicture}
\node[inner sep=0pt] (prod3) {\reducedprod};
\node[below left = 5pt and 50pt of prod3,inner sep=0pt] (prod4) {\reducedprod};
\node[below right = 5pt and 50pt of prod3,inner sep=0pt] (prod5) {\reducedprod};

\node[below left = 5pt of prod4, minimum width = 1.75cm,inner sep=0pt] (intervals) {\shareddomain{U.intervals}};
\node[below right = 5pt of prod4, minimum width = 1.75cm,inner sep=0pt] (set) {\optshareddomain{U.bPowerset}};

\node[below left = 5pt of prod5, minimum width = 1.75cm, inner sep=0pt] (relpoly) {\optshareddomain{U.relPoly}};
\node[below right = 5pt of prod5, minimum width = 1.75cm, inner sep=0pt] (relgrid) {\optshareddomain{U.relGrid}};

\draw (prod3) -- (prod4);
\draw (prod4) -- (intervals);
\draw (prod4) -- (set);
\draw (prod3) -- (prod5);
\draw (prod5) -- (relpoly);
\draw (prod5) -- (relgrid);
\end{tikzpicture}
};

\node[below = 10pt of bottom, rectangle, rounded corners, inner sep = 1pt,scale=.9] {
   \begin{tikzpicture}[font=\footnotesize]
      \node[label=right:Sequence] (c1) {\sequence};
      \node[label=right:Reduced product, below = 10pt of c1] (c2) {\reducedprod};
      \node[right = 100pt of c1, shareddomain, draw, circle, minimum size = 3pt, label=right:Universal] (l1) {};
      \node[pylangdomain, draw, circle, minimum size = 3pt, label=right:Double programs, below = 10pt of l1] (l3) {};
   \end{tikzpicture}
};
    \end{tikzpicture}
  }
  \caption{Date analysis configuration}
  \label{fig:config}
  \vspace{-3em}
\end{wrapfigure}
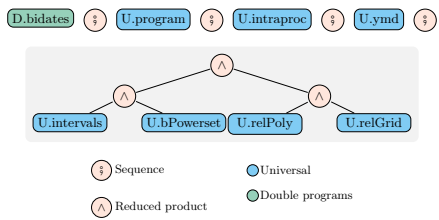
The configuration used by \mopsa{} for our analysis is illustrated in \cref{fig:config}.
The ``D.bidates'' domain corresponds to  
the abstract domain and transfer functions described in \cref{sec:double}.
The ``U.ymd'' domain is the YMD domain (\cref{sec:ymd}).
The last part enclosed in a gray box corresponds to the numerical abstract domain on top of which the YMD domain was built (\cref{sec:num}).

\subsection{Generating counter-example hints}
\label{sec:counter-examples}

We have extended our implementation to provide counter-examples hints when a synchronization assertion cannot be proved safe.
Given our usecase, it is paramount to provide meaningful feedback to users translating law articles into Catala code so they understand why their date computations might be ambiguous (\cref{sec:catala}).
These hints are precise constraints on the considered program that may lead an expression to be rounding-sensitive.
They are especially helpful to provide more precise date ranges for unconstrained dates that may affect rounding sensitivity.
As our approach is incomplete, these hints may be spurious; we however did not encounter this issue in our case study on Catala programs.

This generation of counter-example hints is atypical for static analyses by abstract interpretation.
This approach is permitted here by a simplified setting (variables are assigned once, and the abstract state is partitioned to ensure a high precision) and the use of powerful relational abstract domains.
In a general setting with multiple variable assignments, joins and widenings, most approaches need to perform backward analyses \cite{DBLP:conf/sas/Rival05,DBLP:conf/tacas/AlbarghouthiGC12,DBLP:journals/scp/Mine14}.

This generation of hints works in two steps: it first starts by heuristically selecting the best partition of the abstract state.
The YMD domain may partition the abstract state in order to keep the best precision.
Our heuristic selects the partition with the highest number of desynchronized variables (meaning there has been significant roundings), and the highest number of auxiliary variables for days and months which are constants.
The second step of the hint generation extracts the relevant constraints from the considered abstract state.
This extraction starts by collecting all date variables defined in the program.
For these variables, we evaluate the auxiliary day, month and year variables into intervals, and keep only intervals providing meaningful information (i.e., intervals strictly included in $[1, 31]$ for day variables, strictly included in $[1, 12]$ for month variables, and bounded intervals for year variables).
We then project the relational abstract domain onto the set of auxiliary variables where no meaningful intervals has been extracted to provide linear relations for those.
We show in \cref{fig:mopsaoutput} the exact, unedited output of the hints generated by \mopsa in the case of our running example and highlight their readability.
They correspond exactly to the constraints previously described in \cref{ex:running:rnd}.

\begin{figure}[!t]
  \begin{minted}[escapeinside=||,autogobble,xleftmargin=2em]{text}
  5: assert(sync(current < limit));
                 ^^^^^^^^^^^^^^^
  Desynchronization detected: (current < limit). Hints:
  |$\upa$|month(limit) = 3, |$\upa$|day(limit) = 1, |$\doa$|month(limit) = 2, |$\doa$|day(limit) = 1,
  |$\upa$|month(intermediate) = 3, |$\upa$|day(intermediate) = 1,
  |$\doa$|month(intermediate) = 2, |$\doa$|day(intermediate) = 28,
  month(birthday) = 2, day(birthday) = 29, month(current) = 2, day(current) = [1,29],
  year(birthday) =[4] 0, year(current) = |$\upa$|year(intermediate) = |$\upa$|year(limit)
  = |$\doa$|year(intermediate) = |$\doa$|year(limit) = year(birthday) + 2
\end{minted}
  \caption{\mopsa's output on the running example}
  \label{fig:mopsaoutput}
\end{figure}

\section{Case Study: Application to Catala}
\label{sec:catala}
This section highlights how the results and methods established in the previous section can be applied in the setting of
legal expert systems, and more specifically within Catala~\cite{merigoux2021catala},
a recent domain-specific programming language designed to be understandable by lawyers and
close to the structure of legal texts, with formal semantics that clearly define its behavior to
reduce discrepancies between legal texts and their implementation.

We start by describing rulings and implementations of the law where precise and well-defined date arithmetic is paramount to ensure expected results.
Then, we describe how Catala's implementation of date rounding has recently evolved: from the issues we noticed in Catala's previous off-the-shelf implementation,
to the port to our date calculation library and the introduction of a function-local rounding definition when legal references or interpretations are known, reducing the number of cases where the rounding mode is unspecified.
We finish by explaining the latest implemented feature, which allows the Catala compiler to extract date computations and relies on \mopsa to (dis)prove rounding-insensitivity.

\subsection{Date arithmetic and the law}

Critical software relying on date computations is commonly used by
companies or government agencies to automatically enforce legal dispositions, e.g.,
to check if an application has been filed within the correct time
period, to compute age-related conditions, or to aggregate periods between dates and
compare the result to a fixed duration for eligibility calculation.

In all these cases, there can be heavy financial and legal consequences when a
date computation goes wrong or is subject to diverging interpretation.
In the
case \href{https://supreme.justia.com/cases/federal/us/551/205/}{Bowles v.
  Russell, 551 U.S. 205 (2007)} cited by \citet{bailey2023keynote}, the court gave
Bowles a 17-days notice to file an appeal but this notice was incorrectly
computed from Rule 4(a)(6) and paragraph 2107(c), as it should have been 14
days. When Bowles filed his appeal on the 17\textsuperscript{th} day, the court
system dismissed the appeal on the basis that Bowles should have filed on the
14\textsuperscript{th} day and not trust the notice the court gave him earlier.
In more mundane cases, an incorrect date computation can deprive someone of
their social benefits, or impose a higher late fee than what should be.

These doubts about date computation in software applying the law are all the
more concerning that previous research in code open-sourced by French government
agencies did not show a great deal of transparency or trustful practices on
that particular matter. For instance, the custom programming language M, used by
the French tax authority to compute income tax \cite{mlang}, encodes dates as
mere floating-point numbers where the date is just a decimal number in the
format DDMMYYYY. The French unemployment agency, whose IT system is mostly
implemented in Java, uses a custom date library for its computation
(\texttt{fr.unedic.util.temps.Damj}) but its implementation is omitted from
their only open-source release~~\cite{codesource2018are}.

\subsection{Catala's policy about date rounding}

Recently, the Catala project~\cite{merigoux2021catala,huttner:hal-02936606}
has aimed to bring more
accountability and transparency to programs computing taxes or social benefits
inside government agencies. The Catala language is specifically designed to
allow the easy translation of computational law into code;
in particular, it is based on prioritized default logic~\cite{brewka2000prioritizing},
which enables programmers to closely follow the base case/exception pattern
that permeates the law. To increase confidence and explainability in its
programs, Catala also comes with a formal semantics which is formalized
in the \fstar proof assistant. These formal semantics mostly
focus on Catala's default calculus, the encoding of prioritized default
logic as a programming language, and do not specify all Catala expressions,
including date computations.

Initially, the semantics of the date computations was defined by the behavior of
the \texttt{calendar} OCaml library~\cite{calendar} used inside its interpreter.
However, this library relies on the POSIX behavior which is not always monotonic
and may appear quirky (for instance, it computes \texttt{Jan 31st + 1 month} as
\texttt{March 3rd} for non-leap years) despite its very complete set of features.
These unusual
behaviors prompted a deeper investigation about the corner cases of date
computations and led to the implementation of the library presented in this
paper.
While now integrated in the Catala interpreter, our library is standalone,
and freely available with an open-source license. As the Catala compiler
is implemented in OCaml, so is our library\footnote{Our \fstar formalization can
be extracted to executable but non-idiomatic OCaml code.
In practice, we thus manually reimplement our library in OCaml to use features such
as named arguments or exceptions to provide a more idiomatic API.}, currently packaged with opam;
however, by relying on our semantics, its implementation is straightforward.
We do not foresee any difficulty porting it to other languages, and plan to
do so to support more of the Catala backends, including Python and JavaScript.

The default behavior of our date computation library inside the Catala
interpreter is to raise a runtime exception whenever a date rounding is needed
during a computation. This choice of behavior has been made conservatively
because the decision to round up or down date computations in software enforcing
legal rules is itself a legal rule that has to be specified, as we described 
in the introduction of this paper.
To avoid runtime exceptions, rounding rules can be specified at the
scope level (a precise definition of Catala's scopes is outside the range of this paper,
but it can be considered as a sort of function in Catala) and
should be justified, for example by a legal reference or interpretation.

We applied this methodology to fix the code of the biggest Catala program so
far, which computes the French housing benefits~\cite{merigoux:hal-03933574}.
Articles
\href{https://www.legifrance.gouv.fr/codes/article_lc/LEGIARTI000038814940}{L822-4},
\href{https://www.legifrance.gouv.fr/codes/article_lc/LEGIARTI000038878933}{R823-4}
of the construction and housing Code, as well as article
\href{https://www.legifrance.gouv.fr/codes/article_lc/LEGIARTI000041979743}{L512-3}
of the social security Code, all feature a comparison of the age of the user to
an age constant. However, as the input to the Catala program is not the age of
the user but their birth date, we know such a comparison can be ambiguous if the
user was born on February 29\textsuperscript{th} on a leap year and if the
current date is March 1\textsuperscript{st}. In those situations, we took the
decision to round up the date addition, as shown in \cref{code:ageinferior},
with the \texttt{date rounding increasing} mention. We are currently trying to
contact the relevant government agencies operating the system for clarifications
about how this issue should be handled.

\begin{figure}[!t]
  \begin{minted}[numbers=left,xleftmargin=2em,autogobble,firstline=2,firstnumber=1,lastline=11]{catala_en}
```catala
declaration scope CheckingAgeInferiorEqual:
  input birth_date content date
  input current_date content date
  input target_age content duration # always a number of years
  output age_is_inferior_or_equal_target content boolean

scope CheckingAgeInferiorEqual:
  definition age_is_inferior_or_equal_target equals
    birth_date + target_age <= current_date
  date rounding increasing
```
\end{minted}
\caption{Catala code for checking the age of the user is lower than a constant}\label{code:ageinferior}
\end{figure}

To best benefit the recipient and be in line with the general principle underpinning legal
interpretations of social security law in France, a better solution would be to
perform the computation twice, by rounding up and down, and select the outcome
most favorable to the user in case of disagreement.
The flexibility offered by our library allows us to do that, and we intend to explore
this avenue in future work.
Being able to control precisely where the rounding is done and how is key for developers and
maintainers of such programs, as they are responsible for the legal effect of
the program itself~\cite{typology2022cohubicol}.

\subsection{Detecting potentially ambiguous computations}

Choosing the rounding mode for each date computation allows
us to precisely control the outcome of ambiguous computations. However,
given the pervasiveness of such computations in legal texts, it is also
extremely tedious, and figuring out the cases where an ambiguous
computation could happen is complex. For these reasons, we expect some
developers to delay this
step and wait for incidents to figure out the policy of the
institution operating the program on the matter. But figuring out this policy
might itself be tricky because of the automation frontier~\cite{merigoux:hal-03712130}
strictly separating the developers from the decision-makers in charge of legal
policy decisions.

\begin{figure}[!t]
  \begin{center}
    \resizebox{\textwidth}{!}{
      \tikzstyle{Box2}=[rectangle, draw, rounded corners, fill= white, align= center, copy shadow={draw, shadow xshift=0.5mm, shadow yshift=-0.5mm}]
\tikzstyle{Box}=[draw, text centered]
\begin{tikzpicture}[scale=1,font=\smaller]
  \node (catalaprog) [rectangle, rounded corners, draw, text centered] {\texttt{file.catala}};
  \node (slicing) [Box, right = 1.5em of catalaprog] {Slicing}; 
  \node (dcomp) [Box2,right = 1.5em of slicing] {date-sensitive\\computations}; 
  \node (proggen)  [Box, right = 1.5em of dcomp] {Prog. gen.};
  \node (mopsaprog)  [Box2,right = 1.5em of proggen] {progs.u};
  \node (mopsa) [Box, right = 1.5em of mopsaprog] {\mopsa};
  \node (mopsa_ok) [above right = 1em and 7.5em of mopsaprog] {\faCheckCircle};
  \node (mopsa_ko) [below right = 1em and 6.5em of mopsaprog] {\faWarning + Hints};
  \draw[-stealth,line width=1pt] (catalaprog) -- (slicing);
  \draw[-stealth,line width=1pt] (slicing) -- (dcomp);
  \draw[-stealth,line width=1pt] (dcomp.east)++(0.05,0)  -- (proggen);
  \draw[-stealth,line width=1pt] (proggen) -- (mopsaprog);
  \draw[-stealth,line width=1pt] (mopsaprog.east)++(0.05,0) -- (mopsa);
  \draw[-stealth,line width=1pt] (mopsa.east) -- (mopsa_ok);
  \draw[-stealth,line width=1pt] (mopsa.east) -- (mopsa_ko);
\end{tikzpicture}
    }
  \end{center}
  \caption{Catala date ambiguity analysis pipeline}
  \label{fig:mlang:structure}
\end{figure}
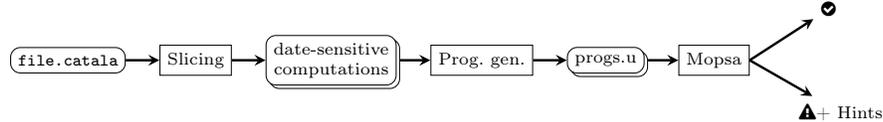

To help developers reach out to the legal services of their institution with
concrete examples of where things can go wrong before production incidents, we
integrated the semantics and abstract domains presented in this paper inside the
ongoing initiative to provide a proof platform for Catala programs
\cite{delaet:hal-03447072}. By connecting the Catala compiler to the \mopsa
static analyzer, we are able to check whether a date computation can be
ambiguous in the context of the program, and often exhibit a counter-example if it is
the case.
We present in \cref{fig:mlang:structure} our analysis pipeline.
It consists of three main phases: program slicing, verification
condition crafting, and analysis -- which may generate counter-examples.

First, we scan the Catala program in one of its intermediate representation and
look for Catala expressions susceptible of raising a runtime exception because of
an ambiguous date computation.
We use classic techniques of program slicing for this
step, selecting only the target sub-expression and then adding the definitions
of variables used in that sub-expression recursively to extract a small, self-contained
program with sufficient information to be analyzed.
This will simplify the counter-example hint generation of \mopsa, which outputs constraints on variables rather than subexpressions of a computation.

Second, we augment the sliced program with the assertions and
other information about its variables that are declared in the original Catala
program to further constrain the search space. So far, our analysis is intraprocedural,
but we are planning to implement an inlining pass to make it inter-procedural.
We then translate the sliced program to \mopsa's toy language
(using the \texttt{.u} extension), which can then be fed
to the static analyzer.

Finally, we run \mopsa on the generated program.
As we have mentioned in \cref{sec:counter-examples}, \mopsa is able to exhibit
potential counter-examples hints.
While these hints are approximate due to incompleteness of the analysis, they are often sufficient to yield real, actionable counter-examples
on the Catala programs that we analyzed.
We extract relevant intervals and linear
constraints and display them to the user, in the format illustrated by \cref{fig:mopsaoutput}.
While the intervals and constraints presented are descriptive,
and sufficient for a programmer to identify concrete counter-examples, they can however
be difficult to grasp for non-experts. Formatting these
constraints in a more readable format is an interesting question, requiring further interaction with lawyers; we leave it as future work.

The implementation of housing benefits in Catala currently consists of about 20,000 lines (including the text of the law directly specifying it) that were written prior to this work.
While automatically analyzing this implementation using our verification pipeline, we found issues in two date computations (one of them being our running example).
In both cases, \mopsa was able to provide actionable counter-example hints.
Several other computations were age computations, which are now handled by a custom scope with a legally interpreted date rounding mode, as shown in~\cref{code:ageinferior}.
Finally, remaining computations rely on durations defined outside of the analyzed scope, which requires an inter-scope analysis in Catala, which is being implemented.
In the meantime, we performed a manual duration extraction in these cases and detected 16 new unsafe (rounding-sensitive) date comparisons, which are real issues.
In all cases, the provided counter-example hints are actionable.
In 10 cases, the issues can only happen with a current date before 2023.
By constraining the year to be greater or equal to 2023, these 10 cases are proved safe.
All date arithmetic programs we have currently extracted or written are small and analyzed within three seconds.

As the number of Catala programs grows, we hope to apply our analyzer at a larger scale, possibly suggesting future avenues for improvement.

\section{Related Work}
\label{sec:rw}
We start by surveying the behavior of mainstream implementations of date arithmetic.
We created a suite of litmus tests involving date-duration additions, and the expected result depending on the rounding mode.
We wrote test drivers for each library, running those tests to decide which rounding mode applies. 

The \texttt{java.time} library \cite{javatime} provides a \texttt{LocalDate} class for dates and a \texttt{Period} class to express durations.
In our tests, the addition is performed by rounding down.
This behavior is explicitly described in the documentation \cite{javaadd}.
To the best of our knowledge, there is no option to use another rounding mode, or fail during ambiguous computations.
In the Python standard library, the \texttt{datetime} module \cite{pydatetime}
provides a \texttt{date} class and \texttt{timedelta} to express durations.
However these durations cannot be defined in terms of months, but only in terms of days.
A third party library called \texttt{dateutil} \cite{pydateutil} provides a replacement feature, \texttt{relativedelta}, able to express durations in months and years.
This library seems widely used, as it ranks within the top 20 most downloaded Python packages.
On our tests, this library rounds down.
This seems to be confirmed by the documentation stating that ``adding one month will never cross the month boundary.''
Similarly to Java, this rounding behavior is not configurable.
The \texttt{boost} C++ \cite{boost} and the \texttt{luxon} \cite{luxon} JavaScript libraries exhibit similar behaviors.

The \texttt{coreutils} implementation of date arithmetic follows a different principle, which is not expressible in our semantics.
When adding months, this implementation first computes an adjusted date which might not be valid. 
This adjusted date $d_a$ is then normalized using POSIX's \texttt{mktime} function.
For example, adding one month to \textsc{2023-03-31} yields adjusted date \textsc{2023-04-31}, which does not exist and is normalized into \textsc{2023-05-01}.
In this case, the behavior is the same as the upper rounding.
There are however cases where its behavior differs: adding one month to \textsc{2023-01-31} yields adjusted date \textsc{2023-02-31}, which is normalized into \textsc{2023-03-03}.
This behavior breaks monotonicity of the addition in the date argument ({\textsc{2023-02-01 + 1 month} is \textsc{2023-03-01}).
In ambiguous computations, the debug mode of the \texttt{date} utility outputs a warning with the following message ``when adding relative months/years, it is recommended to specify the 15th of the months'' -- which is a sufficient condition to avoid any ambiguity.
This semantics is also followed by the \texttt{calendar} \cite{calendar} library of OCaml.

We finish this survey with the case of spreadsheet editors (such as Google Sheets), and highlight an inconsistent behavior we have found in them.
The \texttt{EDATE} function adds a given number of months to a date.
In our experiments, this function silently rounds down.
As such, adding 18 years (that is, 216 months) to \textsc{2004-02-29} yields the date \textsc{2022-02-28}.
These spreadsheets applications also offer the \texttt{DATEDIF} function, which can compute the duration in years between two dates.
In that case, \texttt{DATEDIF}(\textsc{2004-02-29}, \textsc{2022-02-28}) yields 17 years (18 years are reached when the second date is \textsc{2022-03-01}).
This behavior is inconsistent with \texttt{EDATE}.
\citet{DBLP:conf/esop/ChengR15} focus on performing a type analysis of spreadsheet applications, given that a runtime type casting may silently happen and provide unwanted results (similarly to what JavaScript does).
This analysis supports a variety of types, including dates, but as it focuses on type information there is no mention of the value semantics of operations on dates.

The book of \citet{calendrical_calculations} can be seen as the hacker’s delight of calendar computations, with many efficient formulas for day additions, and a wide range of different calendars being presented.
Their work does not mention nor address the issue of month addition, and potential date rounding, which is at the core of our work.
Although we have not needed it for now, we could leverage their approach to optimize the recursive computations of our library.
Similarly, ISO 8601 defines the representation of dates in the Gregorian calendar, but does not address date-duration additions with years or months.

The Formal Vindications start-up developed a mechanized, formally verified implementation of a time management library \cite{BorgesBRRBJ24,formalv} in Coq, computing over dates and time, including specific technical points (timezones, leap seconds).
Their \href{https://gitlab.com/formalv/formalv/-/blob/05b81ab6e6221268d0e0ab3e9d1cd71570473ed3/theories/time/formalTime.v#L784}{duration of a month is defined as 30 days}. Some recent changes allow to round down dates.
A similar effort was developed in Lean 4 by \citet{timelib}, but this library only supports the addition of days to a date.
As a reminder, the Catala project currently targets laws that do not need to go beyond the precision of a day in terms of time management.
Formal Vindications developed a formally verified, high-precision tachograph software for enforcing truck drivers scheduling laws~\cite{DEALMEIDABORGES2021104636}. 

We finish this related work by highlighting similarities between floating-point and date arithmetic.
Floating-point arithmetic is more complex and widely used, but both settings have rounding operators with different modes available.
This similarity has guided us in our search for properties that hold and counter-examples presented in \cref{sec:mechanized}.
The static analysis to prove non-ambiguity of date computations presented in \cref{sec:sa} can be seen as the abstract execution of the computation under both rounding modes, to compare results.
To the best of our knowledge, no such static analysis for floating-point programs try to bound the difference in computations between two rounding modes.
Tools such as Daisy \cite{daisy,daisy_verified}, Fluctuat \cite{fluctuat} and FPTaylor \cite{fptaylor} usually aim at upper-bounding errors between ideal computations over reals and a machine computation using floating-point.

\section{Conclusion and Future Work}
Legal expert systems rely on date computations, which are ambiguously defined in some corner cases.
There are different ways of solving these ambiguities through different rounding operators, where no operator prevails over the others.
We have thus defined semantics for date computations, taking into account these ambiguities to either raise errors, or round the result (either up or down).
This semantics has been implemented into a publicly available OCaml library.
We have studied this semantics and have formally proved several properties they satisfy, and exhibited counter-examples to usual properties they do not satisfy.
We have defined and implemented an analysis that is able to prove an expression to be \textit{rounding-insensitive} in a given program.
This analysis relies on partitioning and relational abstract domains to maintain the best possible precision, and can generate understandable counter-examples hints.
Both our library and the rounding-sensitivity analysis have been integrated within the Catala language -- which focuses on implementing computational laws.
Through our analysis, we found rounding-sensitivity issues in the implementation of the French housing benefits in Catala.
We surveyed the behavior of mainstream date arithmetic libraries, and developed litmus tests that can be used to test new libraries.

There are limitations to our static analysis: its soundness has not been proved mechanically, but the proofs simply lift theorems that have been formally verified.
The current analyzed language is a core imperative language which was sufficient for our case studies.
Having an inter-scope analysis within the Catala to \mopsa translation would improve our precision in the case study.
We plan to craft human-readable error messages from \mopsa's output.
We believe the relevant constraints are already properly extracted by \mopsa and the rest of the work consists in engineering,
in order to inverse the translation from Catala date computations to \mopsa programs.

In spite of these limitations, we believe this paper to be a crucial step into clarifying and 
improving the robustness of many computer programs implementing \enquote{business logic},
often overlooked by formal methods. The widespread use of date arithmetic in programs used 
by companies or government agencies to operate massive financial transfer should 
have prompted a formal analysis of date rounding a long time ago, but the existing literature 
only indicates a recent interest from the formal methods community on the matter.

This work was triggered by the problems we found during interdisciplinary
investigations about French housing benefits using the Catala programming
language. From these investigations surfaced the need for various formal
analysis, which we have thus started integrating into the programming language.
We hope to further develop the integration of static analysis into the Catala
proof platform, thus benefiting both legal and computer science users
by including formal methods advances into development processes of Catala programs.

\subsubsection{Artifact Availability Statement.}
All our development is under open-source licenses, public or in the process of being upstreamed into a public development.
To foster reproducibility of our results, we provide an artefact \cite{date-artefact} containing the formal proofs written in \fstar, our date calculation library, and our ambiguity detection analysis as well as supporting evidence of our case study.

\subsubsection{Acknowledgements.}
We thank the anonymous reviewers for their constructive feedback and support of our work.
We are obliged to Abdelraouf Ouadjaout for making his implementation of partitioning within Mopsa available to us.
We are grateful to David Delmas for the discussions around double semantics, Liane Huttner \& Sarah Lawsky for the interesting discussions around the properties this work targets, and Louis Gesbert for his technical help around the Catala compiler.
We appreciated the many discussions and valuable feedback about this work we got from the whole Catala team.

\bibliography{conferences,cited}

\end{document}